\documentclass[11pt]{article}
\usepackage{nfraconf,graphicx,cite}
\usepackage{times}

\begin{document}

\title{First weak lensing results from the Red-Sequence Cluster Survey}

\author{Henk Hoekstra$^{1,2}$, Howard Yee$^2$, and Mike Gladders$^2$\\
	$^1$ CITA, University of Toronto\\
	$^2$ Dept. of Astronomy and Astrophysics, University of Toronto}

\maketitle

\abstract{The Red-Sequence Cluster Survey (RCS) is a 100 deg$^2$
galaxy cluster survey designed to provide a large sample of optically
selected clusters of galaxies with redshifts $0.1<z<1.4$. The survey
data are also useful for a variety of lensing studies. Several strong
lensing clusters have been discovered so far, and follow up
observations are underway. In these proceedings we present some of the
first results of a weak lensing analysis based on $\sim$ 24 deg$^2$ of
data. We have detected the lensing signal induced by large scale
structure (cosmic shear) at high significance, and find
$\sigma_8=0.79\pm0.08$ (for a CDM cosmology with
$\Omega_m=0.3,~\Omega_\Lambda=0.7,~h=0.7$). Another application of
these data is the study of the average properties of dark matter halos
surrounding galaxies. We study the lensing signal from intermediate
redshift galaxies with $19.5<R_C<21$ using a parameterised mass model
for the galaxy mass distribution. The analysis yields a mass weighted
velocity dispersion of $\langle\sigma^2\rangle^{1/2}= 111\pm5$
km/s. In addition we have constrained for the first time the extent of
dark matter halos, and find a robust upper limit for the truncation
parameter $s<470 h^{-1}$ kpc (99.7\% confidence). A preliminary
analysis also excludes a simple ``prediction'' of lensing according to
Modified Newtonian Dynamics (MOND) at high confidence.}

\section{Introduction}

In these proceedings we present some of the first weak lensing results
based on the $R_C-$band imaging data from the Red-Sequence Cluster
Survey (RCS)\footnote{\tt http://www.astro.utoronto.ca/${\tilde{\
}\!}$gladders/RCS} (e.g., Gladders \& Yee 2000).  The Red-Sequence
Cluster Survey is the largest area, moderately deep imaging survey ever
undertaken on 4m class telescopes. The survey comprises 100 square
degrees of imaging in 2 filters (22 widely separated patches imaged in
$R_C$ and $z'$). Ten patches have been observed using the CFHT 12k
mosaic camera, and the remaining 12 southern patches have been observed
using CTIO. The depth of the survey (2 magnitudes past $M^*$ at $z=1$)
is sufficient to find a large number of galaxy clusters to $z\sim 1.4$.

The survey allows a variety of studies, such as constraining cosmological
parameters from the measurement of the evolution of the number density
of galaxy clusters as a function of mass and redshift, and studies
of the evolution of cluster galaxies, blue fraction, etc. at redshifts
for which very limited data are available at present.

The data are also useful for a range of lensing studies. Strong
lensing by clusters of galaxies allows a detailed study of their core
mass distribution. In addition, given the shallowness of the survey,
the arcs are sufficiently bright to be followed up spectroscopically
(e.g., Gladders, Yee, \& Ellingson 2001). Thanks to the magnification of the
arcs, this allows unprecedented studies of the properties of high
redshift galaxies. Furthermore, in combination with detailed modeling
of the cluster mass distribution, the geometry of the images can be
used to constrain $\Omega_m$.

Here we concentrate on some of the weak lensing applications, for
which we use the $R_C-$band survey data. For the purpose of weak
lensing, we have to date analysed 16.4 deg$^2$ of CFHT and 7.6 deg$^2$
of CTIO data. The data will be used to study the mass distribution of
the galaxy clusters discovered in the survey. A first study of galaxy
biasing as a function of scale is presented in Hoekstra, Yee, \&
Gladders (2001a). In these proceedings we present some early results
of the measurement of the lensing signal by large scale structure
(cosmic shear) and a study of the dark matter properties of (field)
galaxies.  The data analysed so far do not cover complete patches, and
therefore we limit the analysis to the individual pointings.

A detailed description of the data and weak lensing analysis will be
provided in Hoekstra et al. (2001b; 2001c). The results presented in
Hoekstra et al. (2001b) indicate that the object analysis, and the
necessary corrections for observational distortions work well, which
allows us to obtain accurate measurements of the weak lensing signal.

\begin{figure}[b!]
\begin{center}
\hbox{
\includegraphics[width=8cm,bb=40 195 570 700]{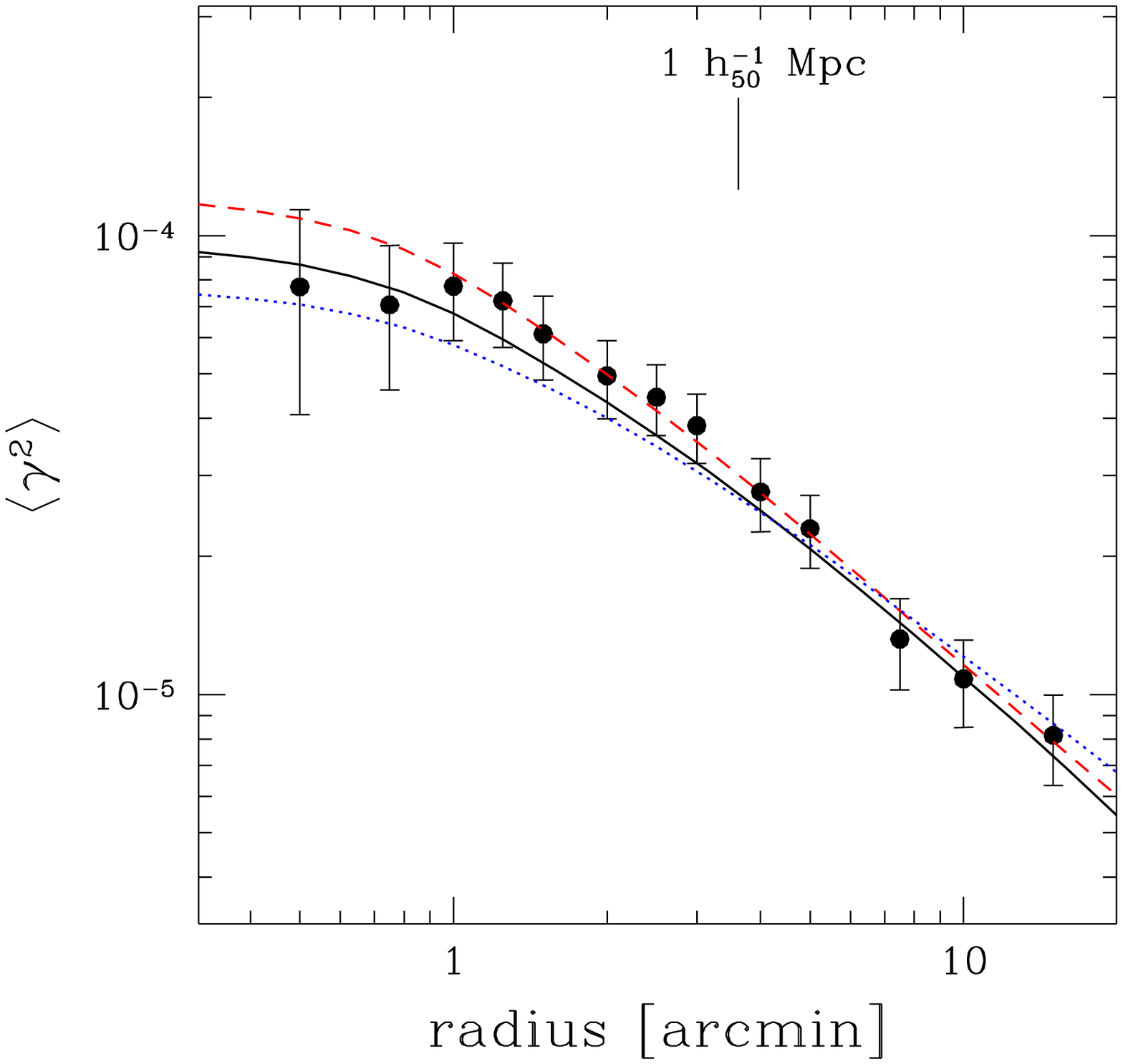}
\includegraphics[height=8cm]{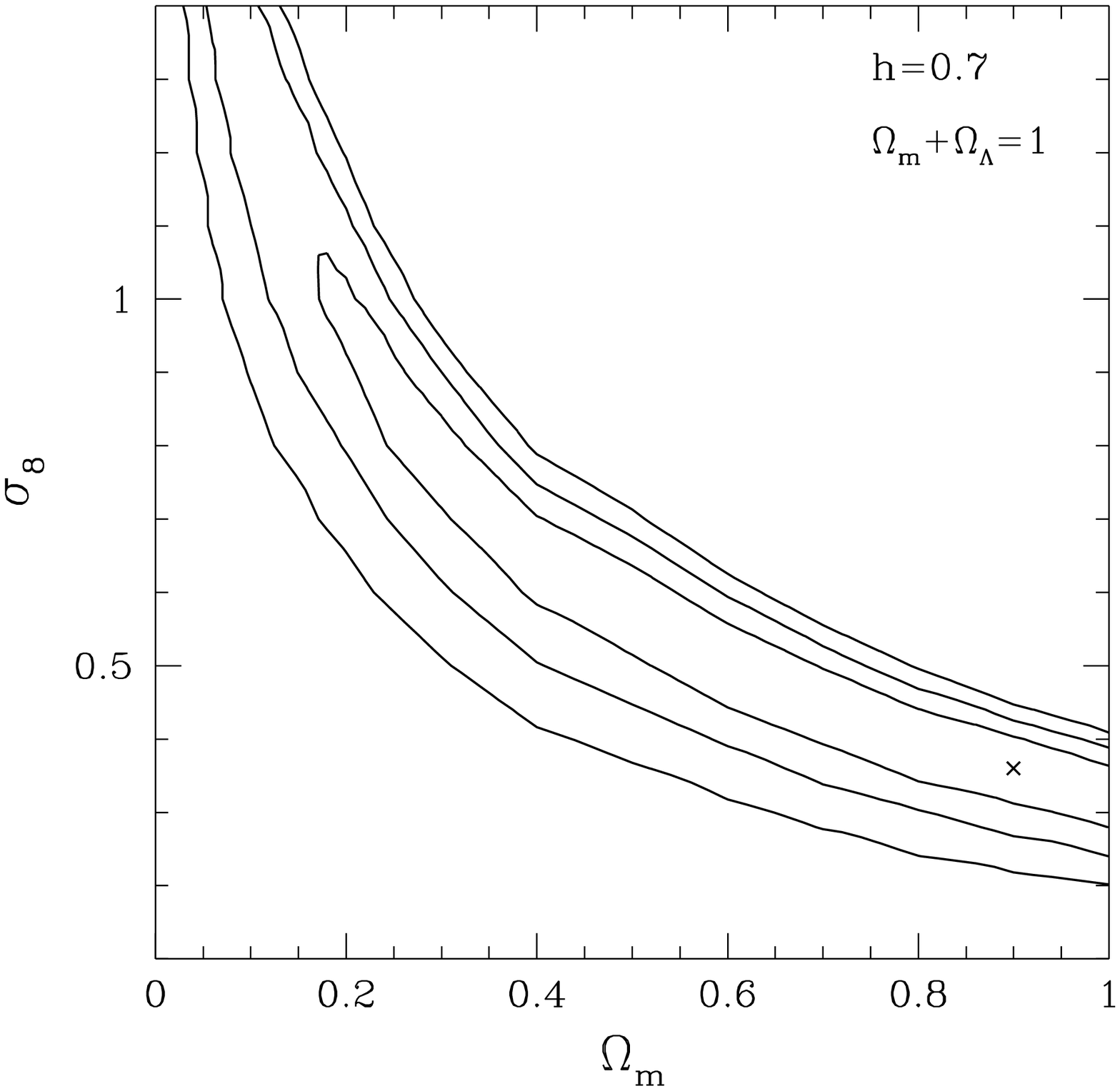}}
\caption{\small {\it left panel}: Measurement of the top-hat smoothed
variance (excess variance caused by lensing by large scale structure)
using galaxies with $20<R_C<24$.  The data consist of 16.4 deg$^2$ of
CFHT data and 7.6 deg$^2$ of CTIO data. The drawn lines correspond to
the expected signals for a SCDM (solid line), OCDM (dashed line), and
$\Lambda$CDM (dotted line) models, using $h=0.7$. The errorbars are
estimated from a large number of realisations of the data set where
the orientations of the galaxies were randomized.  Note that the
points at various scales are strongly correlated.  Under the
assumption that the lensing structures are halfway between the
observer and the sources, a scale of $1~h_{50}^{-1}$ Mpc is indicated.
{\it right panel}: Likelihood contours as a function of
$\Omega_m$ and $\sigma_8$, inferred from the analysis of the top-hat
smoothed variance. The contours have been computed by comparing the
measurements to CDM models with $n=1$, $h=0.7$ and
$\Omega_m+\Omega_\Lambda=1$. The contours indicate the 68.3\%,
95.4\%, and 99.7\% confidence limits on two parameters jointly. Additional
constraints on $\Gamma\approx\Omega_m h$ favour lower values of
$\Omega_m$.
\label{cosmic}}
\end{center}
\end{figure}

\section{Measurement of Cosmic Shear}

The weak distortions of the  images of distant galaxies by intervening
matter  provide  an  important   tool  to  study  the  projected  mass
distribution  in the  universe and  constrain  cosmological parameters
(e.g., van Waerbeke et al. 2001). 

The advantage of our survey, compared to other studies, is that our
data have been acquired using two different telescopes, but have been
analysed uniformly. The results from the two telescopes are in good
agreement, suggesting that the corrections for the various
observational distortions have worked well.

The left panel in Figure~\ref{cosmic} shows the measurement of the
top-hat smoothed variance for the RCS data analysed so far (16.4 deg$^2$
from CFHT and 7.6 deg$^2$ from CTIO). The signal-to-noise ratio of
our measurements is very good, reaching $\sim 6$ at a radius of 2.5
arcminutes.

We use the photometric redshift distribution inferred from the Hubble
Deep Field North and South to compare the observed lensing signal to
CDM predictions. This redshift distribution works well, as was
demonstrated by Hoekstra, Franx \& Kuijken (2000) for the $z=0.83$
cluster of galaxies MS~1054-03. The right panel in Figure~\ref{cosmic}
shows the inferred likelihood contours for a $\Lambda$CDM cosmology,
with $h=0.7$. For an $\Omega_m=0.3$ flat model we obtain
$\sigma_8=0.79\pm0.08$, in excellent agreement with the measurements
of van Waerbeke et al. (2001).

\begin{figure}[b!]
\begin{center}
\hbox{
\includegraphics[height=7.5cm]{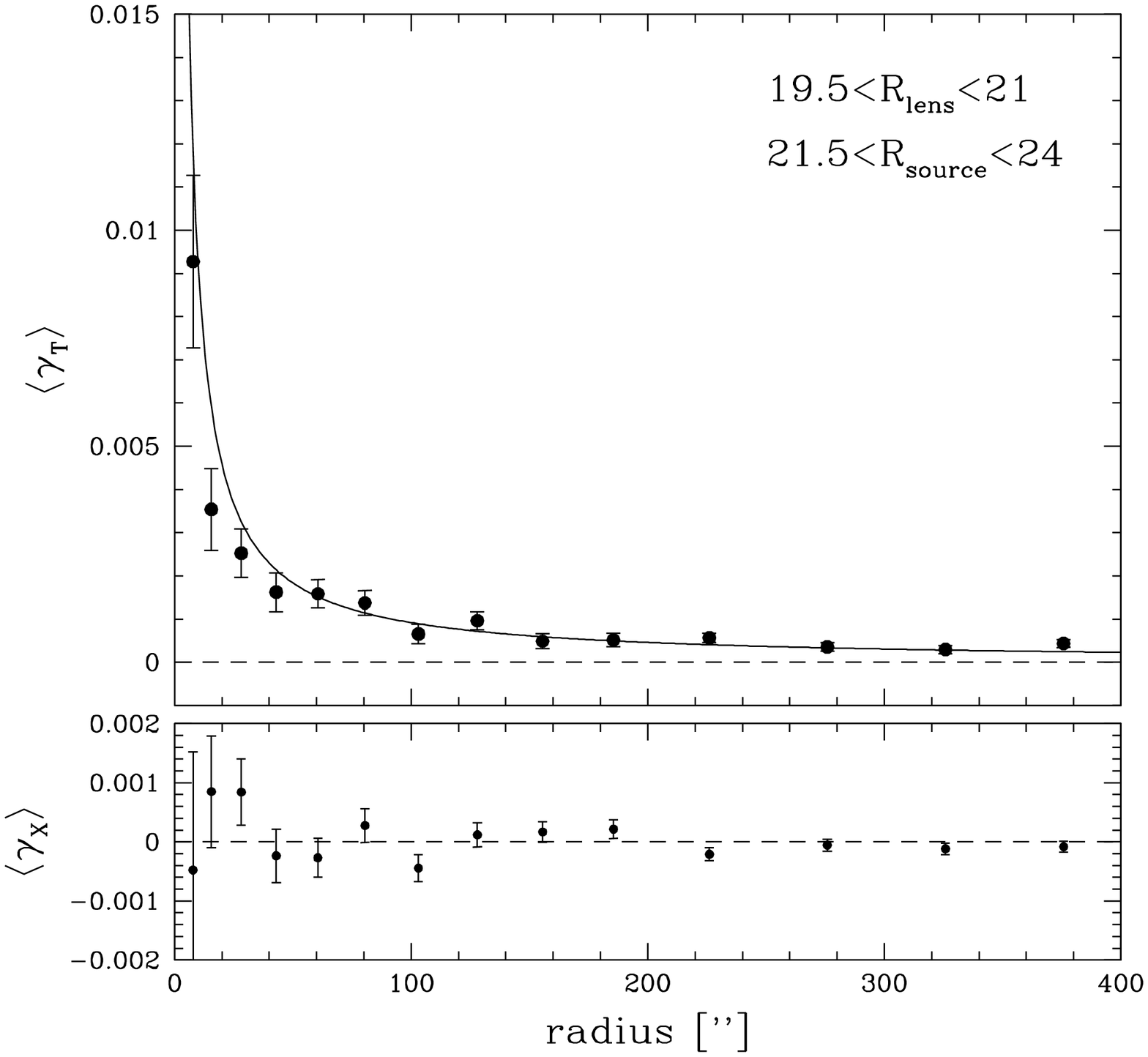}
\includegraphics[height=7.5cm]{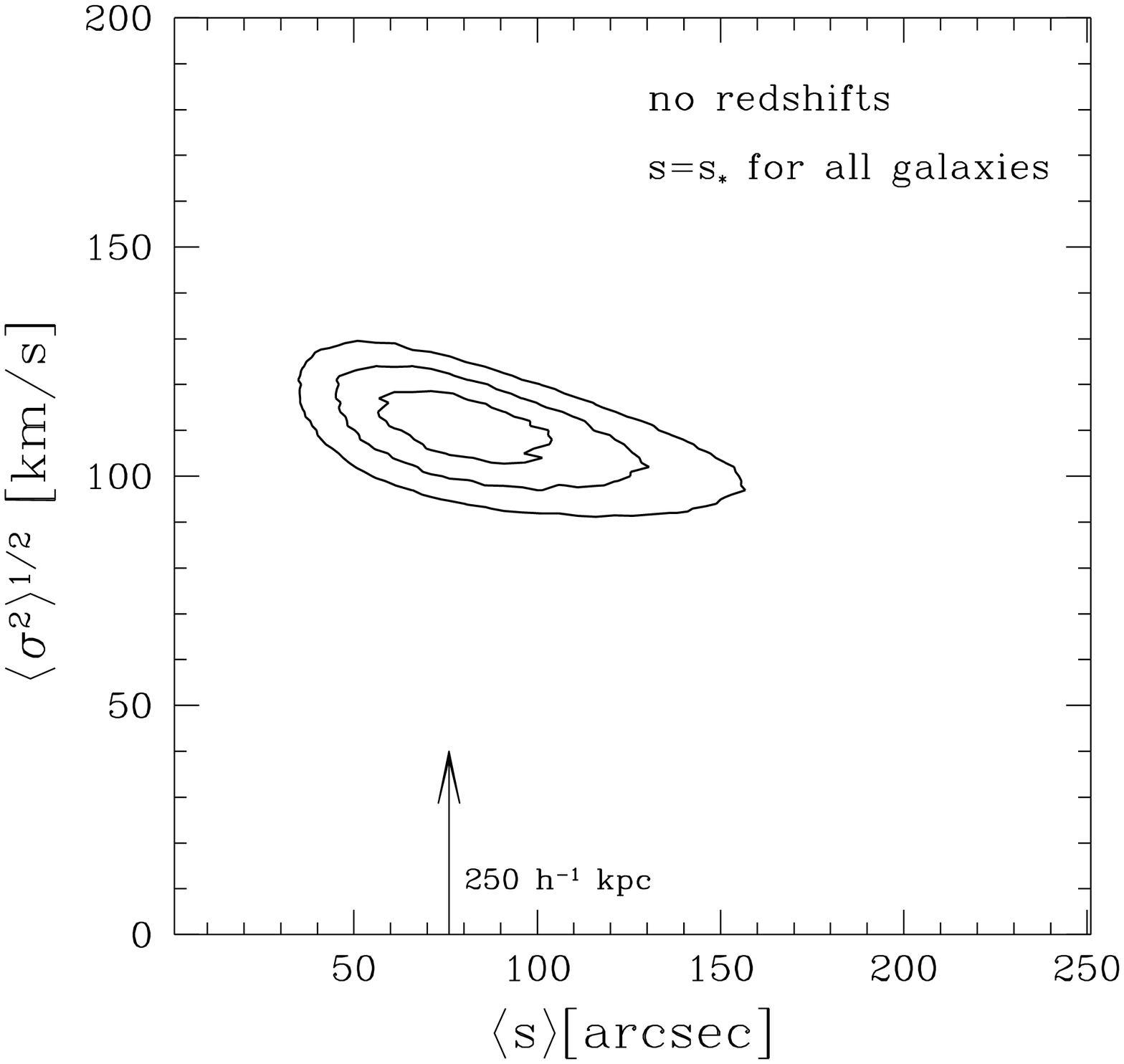}}
\caption{\small {\it left panel}: The ensemble averaged tangential
distortion around galaxies with $19.5<R_C<21$. The solid line
corresponds to the best fit SIS model, for which we find an Einstein
radius $r_E=0.184\pm0.011$ arsec. The lower panel shows the average
signal when the sources are rotated by $\pi/4$. No signal should be
present if the signal in the upper panel is caused by lensing. 
{\it right panel}: Likelihood contours for the mass
weighted velocity dispersion, and the average value of the truncation
parameter $s$. We have also indicated the physical scale when $s$ is
the same for all galaxies. The contours indicate 68.3\%, 95.4\%, and
99.7\% confidence limits on two parameters jointly. For the first
time, we find good constraints on the extent of the dark matter halos
around field galaxies.
\label{galgal}}
\end{center}
\end{figure}

\section{Galaxy-galaxy lensing}

Weak lensing is also an important tool to study the dark matter
halos of field (spiral) galaxies (e.g., Brainerd, Blandford, \& Smail
1996; Fischer et al. 2000). Rotation curves of spiral galaxies
have provided important evidence for the existence of dark matter
halos, but are confined to the inner regions, as are strong lensing
studies. The weak lensing signal, however, can be measured out to
large projected distances, and it provides a powerful probe of
the gravitational potential at large radii. In particular constraints
can be placed on the extent of the dark matter halos.

The lensing signal induced by an individual galaxy is too low
to be detected, and one has to study the ensemble averaged
signal around a large number of lenses.

For the analysis presented here, we use the 16.4 deg$^2$ of analysed
CFHT data, and select a sample of lenses and sources on the basis of
their apparent $R_C$ magnitude. We use galaxies with $19.5<R_C<21$ as
lenses, and galaxies with $21.5<R_C<24$ as sources which are used to
measure the lensing signal. This selection yields a sample of 36226
lenses and $\sim 6\times 10^5$ sources.

The redshift distribution of the lenses is known spectroscopically
from the CNOC2 field galaxy redshift survey (e.g., Yee et al. 2000),
and for the source redshift distribution we again use the photmetric
redshift distribution from the HDF North and South.  The adopted
redshift distributions give a median redshift $z=0.35$ for the lens
galaxies, and $z=0.53$ for the source galaxies.

The ensemble averaged tangential distortion around galaxies with
$19.5<R_C<21$ is presented in the left panel of Figure~\ref{galgal}.
The solid line corresponds to the best fit SIS model, for which we
find an Einstein radius $r_E=0.184\pm0.011$ arsec. 

A better way to study the lensing signal is to compare the predicted
shear (both components) from a parameterised mass model to the
data. We use the truncated halo model from Schneider \& Rix
(1997). The results are presented in the right panel of
Figure~\ref{galgal}. With the adopted redshift distribution we obtain
$\langle\sigma^2\rangle^{1/2}=111\pm5$ km/s. It turns out that the
quoted value is close to that of an $L_*$ galaxy, and our results are
in fair agreement with other estimates. 

In addition, for the first time, the average extent of the dark matter
halo has been measured. Under the assumption that all halos have the
same truncation parameter, we find a 99.7\% confidence upper limit of
$\langle s\rangle <470 h^{-1}$ kpc.  More realistic scaling relations
for $s$ give lower values for the physical scale of $\langle s\rangle$
(Hoekstra, Yee, \& Gladders 2001c), and therefore the result presented
here can be interpreted as a robust upper limit.

We note that the results indicate a steepening of the tangential shear
profile at large radii (which we interpret as a truncation of the halo),
and alternative theories of gravity, such as MOND need to reproduce
this. A plausible description of lensing by MOND was put forward by
Mortlock \& Turner (2001), which suggests that the shear in MOND drops
$\propto 1/r$. A preliminary analysis excludes this prediction at high
confidence.

\end{document}